\documentclass[10pt,conference]{IEEEtran}

\usepackage{graphicx}
\usepackage[dvipsnames]{xcolor}
\usepackage{soul}
\usepackage{cite}
\usepackage{amsmath,amssymb,amsfonts}
\usepackage{algorithmic}
\usepackage{graphicx}
\usepackage{textcomp}
\usepackage{balance}
\usepackage[colorlinks=true,urlcolor=blue,citecolor=black,linkcolor=black]{hyperref}
\usepackage[utf8]{inputenc}
\usepackage[T1]{fontenc}
\usepackage{url}
\usepackage{wrapfig}
\usepackage{caption}
\usepackage{algorithm}
\usepackage{hhline}
\usepackage{fixltx2e}
\usepackage{tabularx}
\usepackage{booktabs}
\usepackage{colortbl}
\usepackage{multirow}
\usepackage{enumitem}
\usepackage{flushend} 
\usepackage{pifont} 
\usepackage[dvipsnames]{xcolor}
\usepackage[scaled=0.86]{helvet}
\usepackage{framed}
\usepackage{orcidlink}
\usepackage{booktabs} 

\newcommand{\RNum}[1]{\uppercase\expandafter{\romannumeral #1\relax}}

\definecolor{light-gray}{gray}{0.9}
\usepackage[framemethod=TikZ]{mdframed}

\mdfdefinestyle{MyFrame}{%
    linecolor=black,
    outerlinewidth=0.15pt,
    roundcorner=3pt,
    innertopmargin=2pt,
    innerbottommargin=2pt,
    innerrightmargin=4pt,
    innerleftmargin=4pt,
    backgroundcolor=light-gray}
    
\mdfdefinestyle{MyFrameSimple}{%
    linecolor=black,
    outerlinewidth=0.15pt,
    roundcorner=3pt,
    innertopmargin=2pt,
    innerbottommargin=2pt,
    innerrightmargin=4pt,
    innerleftmargin=4pt}

\newcommand{\sectopic}[1]{\vspace*{0.1em}\par\noindent{\textit{\bfseries #1}}}

\newcommand {\StepOne}{\color{Red}\ding{202}\color{Black}}
\newcommand {\StepTwo}{\color{Red}\ding{203}\color{Black}}
\newcommand {\StepThree}{\color{Red}\ding{204}\color{Black}}
\newcommand {\SentenceLevel}{\color{MidnightBlue}\ding{202}\color{Black}}
\newcommand {\ParLevel}{\color{MidnightBlue}\ding{203}\color{Black}}

\ifCLASSINFOpdf
\else
\fi

\begin{document}
\title{Rethinking Legal Compliance Automation: Opportunities with Large Language Models}

\author{
\IEEEauthorblockN{Shabnam Hassani \orcidlink{0009-0008-3056-4073}, Mehrdad Sabetzadeh \orcidlink{0000-0002-4711-8319}, Daniel Amyot \orcidlink{0000-0003-2414-1791}}
\IEEEauthorblockA{University of Ottawa\\ Ottawa, Canada\\
Email: \textsf{\{s.hassani, m.sabetzadeh, damyot\}@uottawa.ca}}
\and
\IEEEauthorblockN{Jain Liao}
\IEEEauthorblockA{New Software\\Toronto, Canada \\
Email: \textsf{jian@new.software}}
}
\maketitle

\begin{abstract}
As software-intensive systems face growing pressure to comply with laws and regulations, providing automated support for compliance analysis has become paramount. Despite advances in the Requirements Engineering (RE) community on legal compliance analysis, important obstacles remain in developing accurate and generalizable compliance automation solutions. This paper highlights some observed limitations of current approaches and examines how adopting new automation strategies that leverage Large Language Models (LLMs) can help address these shortcomings and open up fresh opportunities. Specifically, we argue that the examination of (textual) legal artifacts should, first, employ a broader context than sentences, which have widely been used as the units of analysis in past research. Second, the mode of analysis with legal artifacts needs to shift from classification and information extraction to more end-to-end strategies that are not only accurate but also capable of providing explanation and justification. We present a compliance analysis approach designed to address these limitations. We further outline our evaluation plan for the approach and provide preliminary evaluation results based on data processing agreements (DPAs) that must comply with the General Data Protection Regulation (GDPR). Our initial findings suggest that our approach yields substantial accuracy improvements and, at the same time, provides justification for compliance decisions.

\begin{IEEEkeywords}
Legal Compliance, Legal Requirements, Large Language Models, GPT-4, GDPR.
\end{IEEEkeywords}

\end{abstract}

\IEEEpeerreviewmaketitle

\section{Introduction}\label{sec:introduction}

Software-intensive systems are increasingly subject to regulatory mandates. Deriving legal requirements and assessing compliance for these systems requires the ability to accurately interpret and apply legal provisions, including identifying what pertains to software and how. Due to the close relationship between regulatory texts and requirements, the Requirements Engineering (RE) community has extensively studied automated processing of legal texts using Natural Language Processing (NLP) and Machine Learning (ML) to facilitate the definition and analysis of legal requirements. Notable attempts in this direction include works such as~\cite{Breaux2006Towards, Breaux2008Analyzing, Kiyavitskaya2008Automating, Breaux2009Legal, Zeni2015Gaiust, Zeni2016Building, Sannier2017Automated, Sleimi2018Automated, Sleimi2019Query, Amaral2021Ai, Abualhaija2022Automated, Amaral2023ML, Amaral2023NLP,anish2024governance}.
Breaux et al.~\cite{Breaux2006Towards,Breaux2008Analyzing,Breaux2009Legal} extract rights and obligations from regulatory texts. Kiyavitskaya et al.~\cite{Kiyavitskaya2008Automating} 
provide a tool based on rights and obligations to analyze policy documents.
Zeni et al.~\cite{Zeni2015Gaiust,Zeni2016Building} and Sleimi et al.~\cite{Sleimi2018Automated} extract semantic metadata from legal texts. Sleimi et al.~\cite{Sleimi2019Query} further use semantic metadata to build a knowledge base that can be queried.
Sannier et al.~\cite{Sannier2017Automated} automate detection and resolution of cross references.
Amaral et al.~\cite{Amaral2021Ai} extract metadata from privacy policies and automate GDPR compliance checking.
Abulhajia et al.~\cite{Abualhaija2022Automated} automate question answering of compliance requirements. 
\hbox{Amaral et al.~\cite{Amaral2023ML,Amaral2023NLP}} automate completeness checking of DPAs against GDPR. 
In addition, Anish et al.~\cite{anish2024governance} classify security and privacy requirements from obligations in software engineering contracts.

Despite the significant progress made in the RE community for automated legal text processing, important limitations remain. Our goal in this paper is to highlight salient existing limitations and discuss how a rethinking of current automation strategies will help us take advantage of the opportunities provided by \hbox{LLMs~\cite{brown2020language}.}

\subsection{Limitations in Existing Research}
\label{sec:limitations}
\subsubsection{Reliance on sentences as units of analysis}
Existing approaches mostly promote a sentence-by-sentence analysis of legal texts. We have observed three key issues caused by this decision that can significantly affect accuracy.

First, interpreting sentences often requires contextual understanding beyond the immediate sentence structure, as sentences can be related to previously mentioned categories or definitions. A notable example involves sentences with linguistic proforms like ``these''. Consider for example the following snippet from a DPA, introduced in Section~\ref{sec:GDPR}: \textsf{``Depending on the security classification, buildings [...] may be further protected by additional measures. \underline{These} measures include specific access profiles, video surveillance, intruder alarm systems, and biometric access control systems.''} Here, the second sentence alone does not expressly indicate that the mentioned measures are about protecting buildings.

Second, it is common for a legal concept to be defined, with subsequent sentences drawing upon the definition. For instance, in a privacy policy, there may be a statement defining ``personal data''. Subsequent sentences then use ``personal data'' in various contexts. To illustrate, consider the following snippet: \textsf{``By accepting this policy, you are providing \underline{personal data as defined below} [...]: information you provide to us verbally, electronically, or in writing; [...] information obtained from third parties [...]; or information obtained through cookies. Your \underline{personal data} might be disclosed to the tax authorities or other third parties [...]''.} To interpret what may be disclosed as per the provision in the second sentence, one needs to interpret it in view of the definition of ``personal data'' in the first one.

Third, legal texts are frequently interwoven with cross-references~\cite{Sannier2017Automated}. This practice adds complexity by requiring an understanding of the broader content across a collection of legal documents. For instance, consider the following  privacy-policy excerpt: \textsf{``45. (1) A health information custodian may disclose personal health information to [...],
contingent upon the entity meeting the requirements outlined in \underline{subsection (3)}''.} Here, fulfilling the requirements specified in the provision further depends on the content of subsection~(3). The ability to interpret text within the context of cross-references is thus crucial for grasping the full meaning of a given provision.

\subsubsection{Automation strategies lack justification for decisions and are either coarse-grained or entail significant manual effort to build}
Automation for legal texts employs one of the following three strategies or a combination thereof. We argue that these strategies not only fail to offer justification and rationale for decisions -- a capability now achievable with LLMs -- but they either are too coarse-grained, potentially compromising accuracy, or require considerable manual effort, making compliance automation costly to implement.

\textit{\textbf{Strategy 1}}: Extracting metadata or semantic frames and then executing predefined rules over the metadata/frames to check compliance. Examples of works employing this strategy include: (a)~Xiang et al.~\cite{xiang2023policychecker}, who propose a rule- and semantic role-based approach to verify privacy-policy completeness against GDPR; (b)~Amaral et al.~\cite{Amaral2021Model,Amaral2021Ai, Amaral2023ML}, who extract metadata and check for the presence of metadata types envisaged by compliance rules; and (c)~Amaral et al.~\cite{Amaral2023NLP}, who compare GDPR requirements and DPA sentences based on semantic frames for compliance-rule assessment.

\textit{\textbf{Strategy 2}}: Projecting regulatory texts into an embedding space and then measuring the similarity between these texts against (textual) compliance rules or a labelled group of regulatory provisions. An example of research employing this strategy is that of Amaral et al.~\cite{Amaral2021Ai,Amaral2023ML}, who create sentence embeddings and utilize semantic similarity for completeness checking of privacy policies and DPAs.

\textit{\textbf{Strategy 3}}: Using encoder-only single-input transformer models, such as Bidirectional Encoder Representations from Transformers (BERT)~\cite{Devlin2018Bert}, to classify which compliance rules are satisfied by each legal text. An example of work applying this strategy is that of Ilyas et al.~\cite{ilyas2023multi}, who use BERT to construct a multi-label classification pipeline for identifying sentences in DPAs that meet specific compliance rules.

\vspace*{.5em}

\noindent\textit{Limitations.} A notable limitation of all the strategies outlined above is their inability to provide justification for compliance-rule satisfaction or violation. Despite the fact that generative LLMs may still lack sophisticated logical reasoning abilities, they are often capable of offering simple yet valuable explanations to clarify why a specific rule has been met or breached.

In addition to the above general limitation, the three existing strategies suffer from further drawbacks. For Strategy~1, these additional drawbacks are: (1)~There is a significant amount of manual work involved, typically in the form of some kind of qualitative study (e.g., grounded theory) to identify the rules and frames. (2)~There is a challenge when metadata/rules are less prevalent in the dataset and are considered rare instances. In such cases, this strategy requires additional steps such as keyword search~\cite{Amaral2021Ai}. (3)~Regulations are subject to continuous change, which can impact the compliance checking process. This necessitates regularly extracting more concepts or creating new rules, resulting in additional manual effort.

As for Strategy~2, the additional drawbacks include a lack of precision and unsuitability for many rules. As observed by Amaral et al.~\cite{Amaral2023NLP}, semantic similarity is not always indicative of rule applicability due to inherent limitations in capturing the nuances of legal texts. Consider for example the following DPA excerpt (from~\cite{Amaral2023NLP})
: \textsf{``This agreement is between Sefer University [...] the ``Company''; and Levico Accounting GmbH [...]
 WHEREAS
(A)~The Company acts as a Data Controller. 
(B)~Levico Accounting GmbH acts as a Data Processor. (C)~The Processor wishes to lay down their rights and obligations.''} Further, consider compliance rules \textsf{R1} and \textsf{R2}, which we would like to assess for satisfaction: \textsf{R1: ``The DPA shall contain at least one controller's identity and contact details.''; R2: ``The DPA shall contain at least one processor's identity and contact details.''} In this example, although the excerpt satisfies \textsf{R1} and \textsf{R2}, the semantic similarity, as computed by cosine similarity, would be low. This issue can be partially rectified by enhancements such as bi-encoder fine-tuning~\cite{Encoders}. Nevertheless, the coarse-grained semantic similarity measures commonly used in Strategy~2 still tend to overlook the subtleties in legal texts, even after fine-tuning.

Finally, with regard to Strategy~3, the main additional limitation is that this strategy does not offer the flexibility to differentiate between partial and full compliance. To illustrate, consider the following DPA excerpt (borrowed from Amaral et al.~\cite{Amaral2023ML}): 
\textsf{``A description of the nature of the Personal Data Breach, including, if possible, the categories and the approximate number of affected Data Subjects and the categories and the approximate number of affected registrations of personal data.''} Further, consider the following multi-part compliance rule: \textsf{``The notification of personal data breach shall at least include (a)~the nature of personal data breach; (b)~the name and contact details of the data protection officer; (c)~the consequences of the breach; (d)~the measures taken or proposed to mitigate its effects.''}
In this example, the DPA excerpt satisfies some but not all parts of the multi-part rule. In contrast, querying a generative LLM such as ChatGPT would yield more precise answers, such as the following: \textit{``No, the sentence [DPA snippet] does not fully satisfy the rule as it only addresses one part of the required elements.''} Another limitation of Strategy~3 (which is shared with limitations of Strategy~1) is the handling of rare cases. Addressing this issue requires either expanding the dataset, as suggested by Ilyas et al.~\cite{ilyas2023multi}, or disregarding rare cases.

\subsection{Contributions} We present an approach and preliminary results for addressing legal compliance automation in a more end-to-end manner. Our proposed approach has been designed to eliminate the need for explicit specification of metadata, frames, and rules and to provide a finer-grained, semantics-based analysis compared to approaches based solely on embeddings. 

Using LLMs such as Generative Pre-trained Transformer (GPT-3.5)~\cite{OpenAi} and newer models, which support context windows of up to 128k tokens (in contrast to BERT's limited 512 tokens~\cite{Devlin2018Bert}), enables us to consider a much broader context for compliance automation. Furthermore, the generative nature of these LLMs supports prompt-based interactions to provide rationalization and justification for satisfaction or violation of compliance rules. Another key advantage is the few-shot learning capabilities of these newer LLMs, which reduce the need for extensive data labelling and training -- a common requirement in previous research.

To evaluate our proposed approach, we perform a preliminary assessment using DPAs (as previously illustrated in this section and to be explained further in Section~\ref{sec:GDPR}) as the legal artifacts of interest. Our main aim is to examine the accuracy of various generative LLMs in distinguishing compliance and non-compliance. This evaluation focuses on understanding the impact of broader context, transitioning from individual sentences to entire paragraphs, while also considering implications regarding cost and time.

From our experiments with several LLMs, we make the following preliminary observations: (1)~There are considerable variations across different LLMs in terms of accuracy. The choice of LLM is thus likely to be a critical factor when approaching legal compliance automation. (2)~The ability to account for a larger context when interpreting a given regulatory provision leads to substantial improvement gains, as high as 40\% in our experiments.

\textbf{Structure.} Section~\ref{sec:Back} presents background. Section~\ref{sec:related} compares with related work. Section~\ref{sec:approach} describes our approach. Section~\ref{sec:implementation} offers implementation details. Section~\ref{sec:evaluation} plans the approach evaluation. Section~\ref{sec:analysis} outlines our preliminary evaluation results. Section~\ref{sec:conclusion} lays out future research. Section~\ref{sec:package} provides links to our online data and implementation \cite{RE2024Replication}.

\section{Background} \label{sec:Back}
In this section, we provide the necessary legal and technical background for our approach.
\subsection{General Data Protection Regulation (GDPR)}
\label{sec:GDPR}
GDPR~\cite{hoofnagle2019european} is a privacy and data protection law enacted by the European Union to regulate the processing of personal data and enhance individuals' control over their information.
Under GDPR, an organization is identified as either a data controller or a data processor. Data controllers are responsible for determining the purposes and mechanism of collecting and processing personal data, while data processors process the data as per the controller's directives.
In this paper, we use DPAs, as defined by GDPR, to illustrate ideas and conduct a preliminary evaluation. DPAs are binding agreements that specify the responsibilities and rights of controllers and processors. Most of the content in DPAs is software-related, making them of direct interest to the RE community~\cite{Amaral2023ML}. For GDPR compliance, DPAs must follow GDPR Article~28.

\subsection{Large Language Models (LLMs)} \label{subsec:LMs}
We use LLMs to assess the compliance of legal documents. LLMs are statistical models trained to predict and generate coherent, contextually relevant text. These models are broadly categorized into \textit{generative models}, such as GPT~\cite{Radford2018Improving}, which excel in text production, and \textit{discriminative models}, such as BERT~\cite{Devlin2018Bert}, which are optimized for classification and analysis tasks. LLMs undergo extensive pre-training on vast text corpora, enabling them to grasp language nuances, contextual information, and complex linguistic patterns. 

Users can guide LLMs by providing \textit{prompts}, which are specific instructions or queries that guide the model to generate relevant and contextually appropriate responses~\cite{hariri2023unlocking}. Various prompting strategies exist for LLMs~\cite{Liu2023pre}, such as zero-shot, few-shot, chain-of-thought, and tree-of-thought prompting. In this paper, we explore zero-shot prompting, where the LLM generates responses without prior specific examples, relying solely on the task's context and instructions.

To optimize their performance for specific tasks, such as compliance checking, LLMs are \textit{fine-tuned} -- a process that involves adjusting the pre-trained models to suit particular domain requirements~\cite{Vaswani2017Attention}.
For example, fine-tuning an LLM on GDPR-compliant documents can enhance the LLM's accuracy in interpreting and applying GDPR provisions.

Our preliminary experimentation in this paper encompasses three LLMs: BERT~\cite{Devlin2018Bert}, GPT~\cite{Radford2018Improving}, and Mixtral~\cite{Mixtral}. BERT uses bidirectional training of transformers -- a popular attention model for language modelling. BERT comes in two sizes: BERT base and BERT large. BERT large can be more accurate for certain language understanding tasks but is computationally more expensive. BERT can differentiate between capitalized and non-capitalized text. With BERT uncased, the text is first converted to lowercase before tokenization, whereas with BERT cased, the tokenized text remains the same as the input text with respect to capitalization. Previous RE research favours the cased model for analyzing requirements specification and legal documents. For our experimentation with BERT, we utilize BERT base cased. GPT is a transformer-based family of models, pretrained to predict the next token in a sequence. 
GPT has shown the ability to capture subtle linguistic patterns and excel in many language understanding and generation tasks. We use GPT-3.5-turbo and GPT-4-turbo, more specifically gpt-3.5-turbo-0125 and gpt-4-0125-preview, for our experiments. 
Mixtral is a specialized LLM that integrates multiple linguistic capabilities, including translation, summarization, and question answering. It employs a hybrid architecture, combining traditional rule-based approaches with ML to achieve accurate language understanding and generation. We use Mixtral-8x7B-Instruct-v0.1 for experimentation. 
\section{Related Work} \label{sec:related}
Significant progress has been made in completeness and compliance checking of legal requirements and software-related regulatory artifacts through the application of traditional machine learning algorithms and LLMs. In this section, we compare our approach with existing methods, emphasizing differences in methodology and the limitations of prior work.






Among regulatory artifacts relevant to software, privacy policies have arguably received the most attention. Traditional ML-based approaches for completeness and compliance checking of privacy policies have largely relied on datasets or conceptual frameworks proposed by Wilson et al.~\cite{wilson2016creation}, Torre et al.~\cite{torre2019using}, Liu et al.~\cite{liu2021have}, and Amaral et al.~\cite{Amaral2021Model}.

Some previous studies link privacy policies to software code. Fan et al.~\cite{Fan2020Empirical} use traditional ML classifiers to evaluate GDPR compliance in mobile health applications, while Hamdani et al.~\cite{Hamdani2021Combined} combine ML and rule-based methods for automated compliance checking of privacy policies with GDPR. Xie et al.~\cite{Xie2022Scrutinizing} employ Bayesian classifiers and NLP to examine the compliance of privacy policies in virtual personal assistant applications. Amaral et al.~\cite{Amaral2021Ai} combine NLP and feature-based learning to extract metadata from privacy policies and automate GDPR compliance checking. These works focus exclusively on privacy policies. Furthermore, in terms of an automation strategy, they pursue one or a combination of the following: (1)~metadata extraction, (2)~semantic-similarity comparison, and (3)~hand-crafted rules; these strategies pose problems as discussed in Section~\ref{sec:limitations} under Strategies~1 and~2.

Harkous et al.~\cite{harkous2018polisis}, Mousavi et al.~\cite{mousavi2020establishing}, and Tang et al.~\cite{tang2023policygpt} respectively use Convolutional Neural Networks, BERT, and GPT for privacy-policy analysis. Aside from, again, focusing exclusively on privacy policies, these works remain premised on metadata extraction, thus suffering from the limitation discussed under Strategy 1 in Section~\ref{sec:limitations}. In the case of Mousavi et al.~\cite{mousavi2020establishing} and Tang et al.~\cite{tang2023policygpt}, which are respectively using BERT and GPT, we observe that both use LLMs in a discriminative mode. BERT lacks a decoder-based architecture that can be prompted for generative tasks. As for Tang et al.~\cite{tang2023policygpt}, they do not exploit the generative capabilities of GPT and instead choose to use it for classification in a way similar to more traditional approaches.

Another software-related regulatory artifact that has been studied more recently are DPAs (introduced in Section~\ref{sec:GDPR}). Amaral et al.~\cite{Amaral2023ML} combine traditional ML classifiers with cosine similarity-based classification to assess the completeness of DPAs against GDPR. This involves checking each sentence embedding from DPA against all embeddings for all rules. 
More recently, Amaral et al.~\cite{Amaral2023NLP} have introduced an NLP-based approach using semantic frames to check DPA compliance with GDPR. Semantic frames are extracted from both the sentences in the DPAs and the GDPR rules, followed by a comparison to identify similarities and discrepancies between the two sets of frames. 
These works (1)~are sentence-level, (2)~rely on metadata extraction, and (3)~compare semantic similarity, or use semantic frames and rule-based methods that have issues related to Strategies~1 and~2. In contrast, our approach exploits LLM capabilities, aiming to reduce the dependency on feature-based learning and semantic similarity.

Ilyas et al.~\cite{ilyas2023multi} explore multiple AI solutions
for DPA compliance checking, including traditional ML classifiers, Bidirectional Long Short-Term Memory (BiLSTM), and BERT. This work has been applied to a specific set of rules, and excludes several regulatory rules such as metadata and controller rights which earlier work by Amaral et al.~\cite{Amaral2023ML} addresses, most likely due to lack of sufficient data or the recognition that these requirements are difficult to analyze when the context window is limited to individual sentences. This approach overlooks the broader textual context of the input sentences from the input text and treats rules merely as labels. 
In contrast, our proposed approach utilizes prompting strategies made possible by generative LLMs, enabling us to reason about compliance within a context broader than individual sentences and also alleviating the need to manually encode compliance requirements into formally specified rules.

The idea of broadening the context for the analysis of legal provisions can, in principle, be generalized beyond paragraphs. For example, Sun et al.~\cite{sun2023pearl} and Chen et al.~\cite{chen2023walking} propose approaches for handling long documents. The former approach works by decomposing queries into a sequence of actionable tasks, structuring interaction with the document through a systematic plan and execution process. The latter approach creates a summary-based tree structure from long texts, enabling the model to navigate and retrieve information efficiently in response to specific queries. These works raise the prospect that legal documents can be analyzed in their entirety for legal compliance. Unfortunately, as of this writing, these techniques fall short in the legal domain, where it is essential to preserve the details of the regulatory text and fully grasp the legal implications. The decomposition of queries into actionable tasks as proposed by Sun et al.~\cite{sun2023pearl} could lead to an oversimplification of legal texts, thereby compromising the integrity and depth of the interpretation necessary for effective compliance checking. In a similar manner, the summary-based approach of Chen et al.~\cite{chen2023walking} risks losing essential legal nuances, noting that legal documents feature interconnected information, where precision of wording and context is paramount.

Finally, we note that our research is not the first to recognize the potential of LLMs for compliance automation and inconsistency/violation detection. Berger et al.~\cite{berger2023towards} propose a combination of retrieval and LLM-based zero-shot learning to automate compliance checking of audit decisions, while Fantechi et al.~\cite{fantechi2023inconsistency} introduce an LLM-based approach for identifying inconsistencies in natural-language requirements. Although this latter approach does not explicitly address legal compliance, we share some common elements with it. The main distinction between our research and the recent studies mentioned above lies in our interest in creating appropriate units of analysis for legal automation. In doing so, our research aims to explore the inclusion of cross-referenced content as a way to assemble a self-contained context, with just the right amount of information to support answering specific questions about compliance and non-compliance. 
\section{approach} \label{sec:approach}
\begin{figure}[!t]
    \includegraphics[width=.95\linewidth]{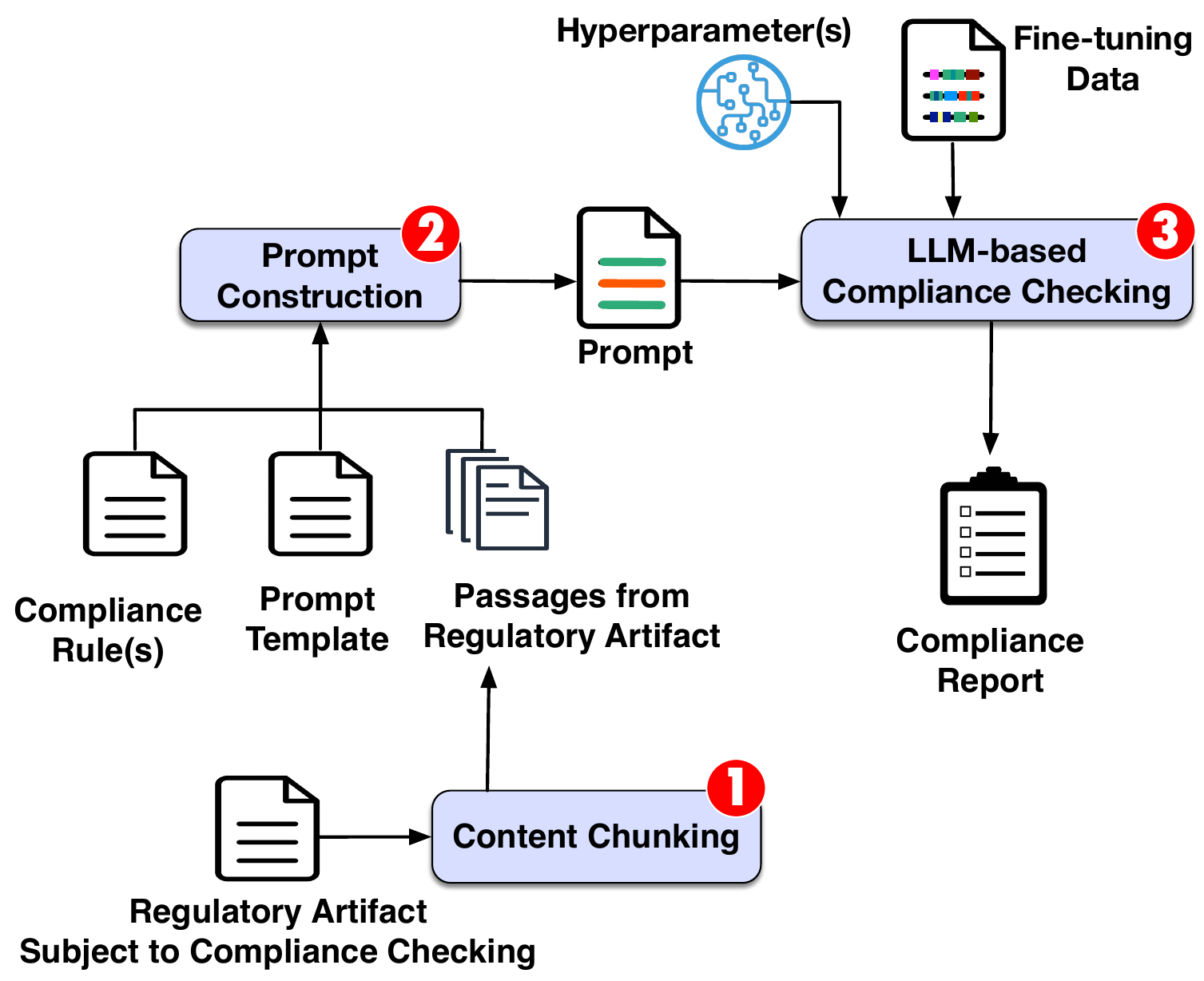}

\caption{Approach overview.}
\label{fig:approach}
\end{figure}

Figure~\ref{fig:approach} presents an overview of our approach, which uses a systematic method employing LLMs to evaluate regulatory artifacts for compliance with specific regulations. 
The approach consists of three steps. The first step (\StepOne) is to create passages from the regulatory artifact that the user wants to verify for compliance, e.g., a DPA. The second step (\StepTwo) is to generate a prompt for LLMs. This step takes as input the passages generated in the first step, along with the compliance rules to be verified (e.g., from the GDPR), and a configurable prompt template. The third step (\StepThree) is to present the prompt generated in the second step to LLMs to draw inferences about compliance and non-compliance. The output of the approach is a compliance report, accompanied by an explanation and justification for each determination.

\begin{figure*}[!t]
\centering\mbox{\hspace*{0em}}
    \includegraphics[width=\linewidth]{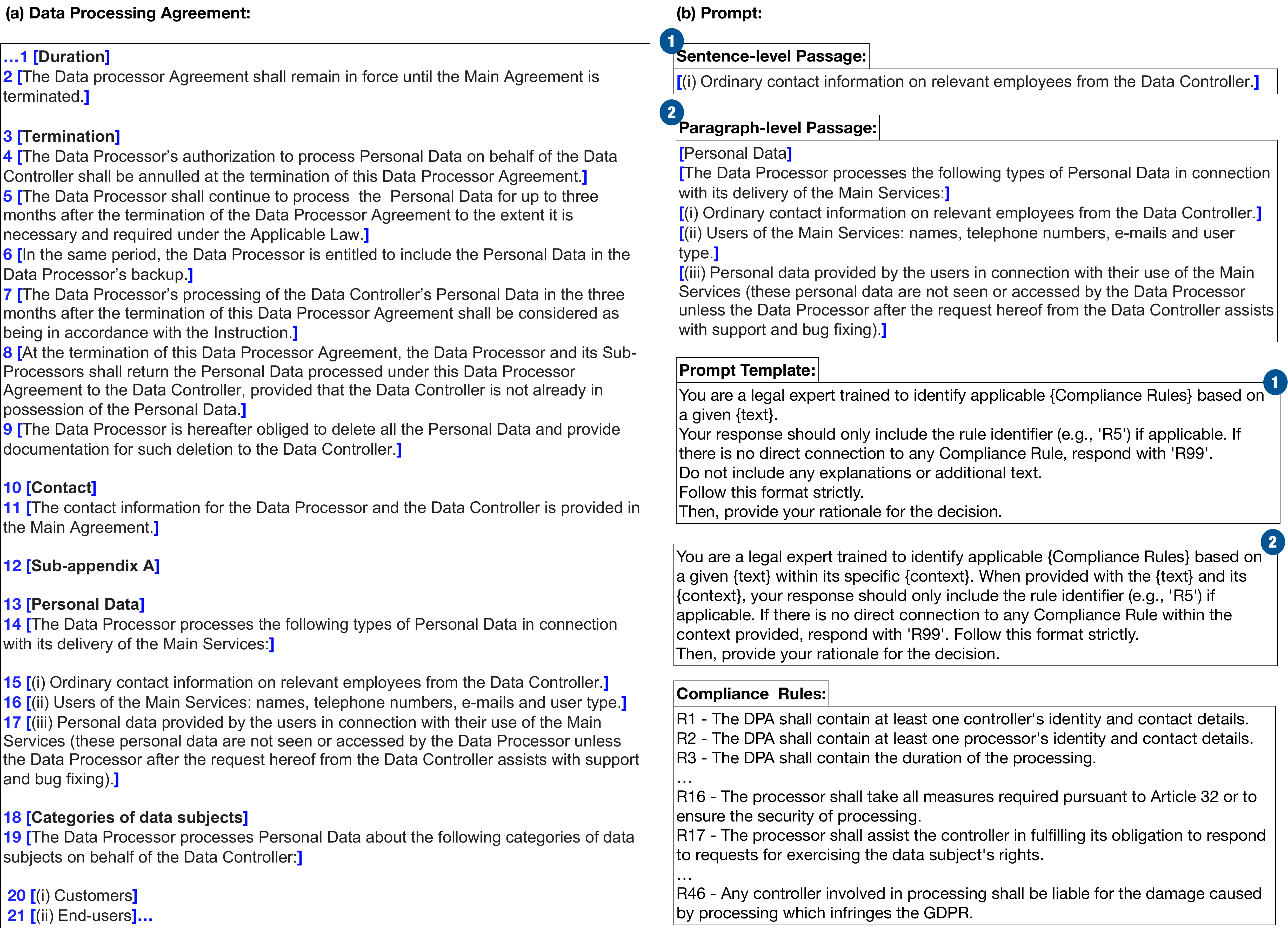}

\caption{(a)~Illustrative data processing agreement (DPA), (b)~Prompt including three parts: Passages from Regulatory Artifact (sentence-level (\SentenceLevel) or paragraph-level (\ParLevel) input), Prompt Template, and Compliance Rules.}
\label{fig:example}
\vspace*{-1em}
\end{figure*}

\sectopic{Step~1)~Content Chunking.}
The regulatory artifact subject to compliance checking is fed to the pipeline. The pipeline segments the regulatory artifact into manageable content chunks,
primarily paragraphs (Fig.~\ref{fig:approach}, Step~\StepOne).
This process ensures that the chunks, once incorporated into the overall prompt, will fit within the token limit of the LLMs used. 
If a full paragraph were to exceed the token limit, it would need to be either truncated or summarized. However, in our investigation of LLMs, we did not encounter such situations due to their reasonably large token limit.
The output of this step is a set of passages each within the LLM's token limit, ready for prompt construction.
Figure~\ref{fig:example} shows a snippet of a DPA as the regulatory artifact subject to GDPR (from Amaral et al.~\cite{Amaral2023ML}), along with examples of passages for both the sentence level (\SentenceLevel) and the paragraph level (\ParLevel).

\begin{figure}[!t]
\centering\mbox{\hspace*{0em}} \includegraphics[width=1\linewidth]{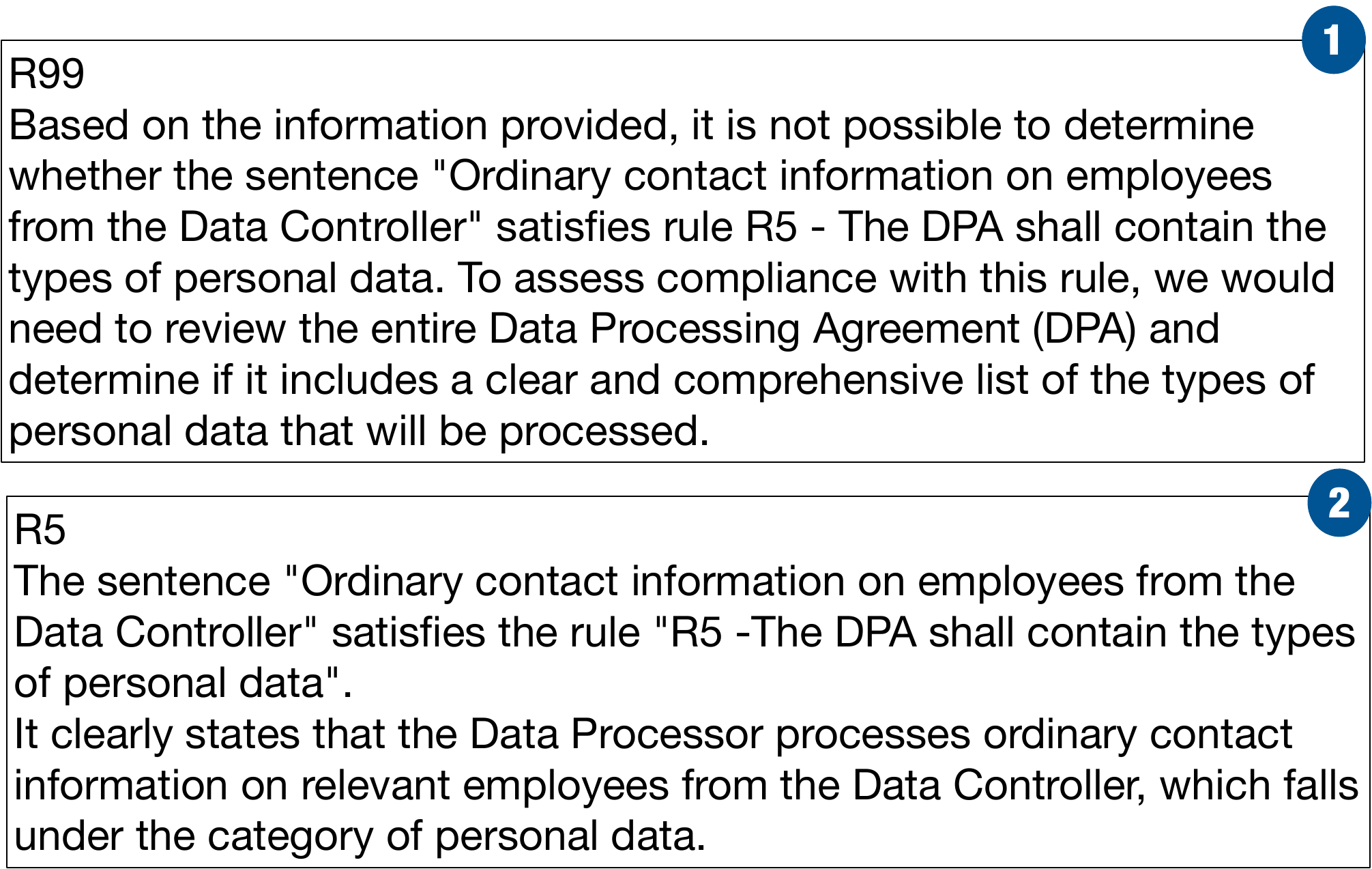}
\vspace*{.5em}
\caption{Illustrative Compliance Report generated by GPT-4 Turbo: for sentence-level inputs (\SentenceLevel) without context, and for paragraph-level inputs (\ParLevel) with context.}
\label{fig:example-report}
\end{figure}
    
\sectopic{Step~2)~Prompt Construction.}
Our approach constructs tailored prompts to guide the LLMs in eliciting rule-specific responses, based on the input compliance rules. 
Prompts, designated to extract both the rule applicability and the LLMs' explanation and justification, consist of three parts: \emph{Prompt Template}, \emph{Compliance Rules}, and \emph{Passages from Regulatory Artifact} (Fig.~\ref{fig:approach}, Step~\StepTwo).

For instance, Fig.~\ref{fig:example} (bottom-right) shows a (trimmed) list of 46 regulatory rules for evaluating DPAs. We denote rules as $R_i$ along with their definitions. In our example, $i$ ranges from 1 to 46, with 99 being a sentinel value indicating non-applicability.
In Fig.~\ref{fig:example}, we illustrate that for sentence-level analysis (\SentenceLevel), the Prompt Template differs slightly from that used for paragraph-level analysis (\ParLevel). Specifically, in sentence-level analysis, the model is provided with a single sentence and asked to make predictions based solely on that sentence. In contrast, in paragraph-level analysis, the model is tasked with making inferences over the input text within the context of the entire paragraph. This difference enhances the model's ability to effectively infer meaning based on the given passage.

The prompt is presented in a structured chat format for the LLMs, specifying roles for system instructions, user inquiries, and the assistant's responses, following best practices where applicable~\cite{OpenAiFinetuning,Chatcompletion}.
The system role provides instructions for the model to follow (detailed in the Prompt Template in Fig.~\ref{fig:example}). The user role presents input that the model should respond to (detailed in the Passages from Regulatory Artifact in Fig.~\ref{fig:example}). The assistant role represents the model's response to the user's input (detailed in the Compliance Report in Fig.~\ref{fig:example-report}).

Additional examples demonstrating different input texts at both the sentence and paragraph levels, along with the model's explanation and justification, are available in our online repository~\cite{RE2024Replication}. The corresponding implementation (see Section~\ref{sec:implementation}) is also provided in the code section of the repository~\cite{Code}.
\sectopic{Step~3)~LLM-based Compliance Checking.}
Once tailored prompts are constructed, we employ zero-shot learning, where the model generates responses (Fig.~\ref{fig:example-report}) without prior training on similar tasks, based solely on the instructions provided in the prompt (Fig.~\ref{fig:example}).
In the future, this process may be enhanced with (optional) fine-tuning to further refine the models' responses to the specific language of DPAs. 

Subsequently, the generated responses are analyzed to determine the compliance of the DPA text with the GDPR, identifying areas of compliance and non-compliance in a comprehensive report (Fig.~\ref{fig:approach}, Step~\StepThree).
An example of the model's explanation and justification is provided in Fig.~\ref {fig:example-report},
for the input text without context (\SentenceLevel), and for the case where the paragraph-level context is provided (\ParLevel). 
\section{Implementation}\label{sec:implementation}
Our approach, described in Section~\ref{sec:approach}, has been implemented in Python. We publicly release our implementation~\cite{RE2024Replication} to encourage future research and facilitate replication. The state-of-the-art LLMs that we experiment with are: \emph{Phi-2}~\cite{Phi-2}, \emph{Mistral-7B}~\cite{Mistral}, \emph{Mistral-7B-Instruct}~\cite{MistralInstruct}, \emph{Zephyr-7B}~\cite{Zephyr}, \emph{Mixtral-8x7B-Instruct-v0.1}~\cite{Mixtral}, all from Hugging Face~\cite{HuggingFace}, using the Transformers 4.35.2 library operated in PyTorch 1.10.2+cu113, as well as \textit{gpt-3.5-turbo-0125} and \textit{gpt-4-0125-preview} from OpenAI~\cite{OpenAi} through the OpenAI Python package. We further implement one baseline approach for comparison: \emph{bert-base-cased} from Hugging Face~\cite{HuggingFace}. To quantize and further fine-tune the LLMs, we use the Transformers 4.35.2 library, Peft library version 1.5.3, and the trl library version 0.7.10.


\section{Plan for the Evaluation} \label{sec:evaluation}
Our proposed approach is designed to be applicable across a variety of legal documents. Here, we instantiate it for DPAs, focusing on determining the effectiveness of LLMs in identifying GDPR compliance. DPAs are chosen for their complex and detailed nature, providing a rich basis for evaluation within the RE community~\cite{Amaral2021Model, Amaral2023ML}.

\subsection{Research Questions}

\textbf{RQ1: How do state-of-the-art LLMs, both open-source and closed-source, fare against one another in terms of accuracy for zero-shot learning and fine-tuning in regulatory compliance tasks?}
This research question compares generative LLMs in terms of their ability to understand and apply regulatory requirements.
To answer RQ1, we use \emph{Accuracy}, \emph{Precision}, \emph{Recall}, and \emph{F-score} metrics. 

\textbf{RQ2: Compared to traditional sentence-level analysis, how does incorporating paragraph context and the textual specification of compliance rules enhance the performance of  compliance checking?}
This research question aims to assess the enhancement in accuracy of compliance checking brought about by the integration of paragraph-level context and rules, as opposed to traditional single-input sentence-level methods. To answer RQ2, we report the performance using \emph{Accuracy}, \emph{Precision}, \emph{Recall}, and \emph{F-score} metrics.

\textbf{RQ3: What are the cost and time implications associated with our proposed approach?}
RQ3 provides a practical assessment of the approach's resource efficiency, considering time and costs in terms of computing and financial resources.

\subsection{Dataset}\label{subsec:data}
Our evaluation uses a dataset of DPAs, borrowed from Amaral et al.~\cite{Amaral2023ML}, which has been manually annotated with GDPR compliance decisions. This dataset, which we subjected to further manual preprocessing to ensure data quality, has been built at the sentence level. This dataset is partitioned into training ($P$) and evaluation ($E$) sets.



\subsection{Experiments}
This section describes the experimental setup and procedures designed to answer our research questions.
\sectopic{Experiment~\RNum{1}.} This experiment answers RQ1. Following Step~\StepOne{} of our approach (Section~\ref{sec:approach}), for every document $p\in P$
from~\cite{Amaral2023ML}, we do data preparation through content chunking.
The input DPAs are segmented into sentences and paragraphs (passages) to incorporate paragraph-level data.

Next, prompt templates, compliance rules, and data chunks are paired to construct the prompt as per Step~\StepTwo{} of our approach, with prompt details further elaborated in Fig.~\ref{fig:example}.
The selected LLMs 
are configured with the temperature hyperparameter set to 0.2. This minimizes variation in responses, thereby making outputs more deterministic while maintaining the ability to generate diverse responses.
For each LLM, we conduct zero-shot experimentation on both sentence-level and paragraph-level passages (Step~\StepThree{} of our approach). The performance is measured against a gold standard set~\cite{Amaral2023ML} using the evaluation metrics listed in Section~\ref{subsec:metrics}.


Optionally, fine-tuning is conducted on the best-performing models from the zero-shot experiments. Fine-tuning is likely to better guide the models in following instructions by exposing them to a set of examples. Furthermore, in relation to Step \StepTwo{} of our approach, we will likely explore several alternative prompt strategies to determine which strategy leads to the best decision verdicts, explanations, and justifications.



\sectopic{Experiment \RNum{2}.} This experiment answers RQ2. We compare the best configuration of our approach resulting from EXPI with single-input models like BERT~\cite{Devlin2018Bert}, which process sentences in isolation, lacking broader context and explicit rule definitions; see Section~\ref{subsec:metrics} for the evaluation metrics used.

\sectopic{Experiment \RNum{3}.} This experiment answers RQ3 by measuring the execution time and cost of our approach from the perspective of end-users to ensure its feasibility for real-world applications. We use a modest platform to replicate the resources available to end-users. Specifically, we use the free version of Google Colab Cloud with the following specifications: Intel Xeon CPU@2.30GHz, Tesla T4 GPU, and 13GB RAM. 
To conduct the GPT experiments, we utilize OpenAI's token-based plan (paid subscription service). 

\subsection{Metrics}\label{subsec:metrics}
For each input passage (sentence of paragraph) $m$ from $E$ (as defined in Section~\ref{subsec:data}), we have the model predict whether $m$ satisfies each compliance rule $R$. We evaluate the quality of the predictions using \emph{Precision}, \emph{Recall}, \emph{F-score}, and \emph{Accuracy}, according to their standard definitions.

A true positive (TP) arises when the model correctly predicts $R_j$ as ``satisfied'', and a true negative (TN) when it correctly predicts $R_j$ as ``not satisfied''. A false positive (FP) arises when $R_j$ is incorrectly predicted as ``satisfied'', while a false negative (FN) arises when $R_j$ is incorrectly predicted as ``not satisfied''.


\section{Preliminary Results}\label{sec:analysis}

\subsection{Answers to Research Questions}\label{subsec:result}
Below, we provide initial answers to our research questions, drawing from our ongoing experimentation and following the analysis procedures detailed in Section~\ref{sec:evaluation}.

\textit{\textbf{RQ1.}} 
Our experiments with \emph{Phi-2}\cite{Phi-2}, \emph{Mistral-7B}\cite{Mistral}, \emph{Mistral-7B-Instruct}\cite{MistralInstruct}, and \emph{Zephyr-7B}\cite{Zephyr} indicate that these LLMs fare poorly as alternatives for our intended purposes. There are multiple cases where these LLMs do not follow predefined instructions in the prompt. Even when fine-tuned with sentence-level passages, these models did not exhibit the desired behaviour. We note that, in view of Jiang et al.'s findings \cite{jiang2023mistral}, indicating that \emph{Mistral-7B} \cite{Mistral} outperforms both \emph{LLaMA2-7B} \cite{Llama-2-7b} and \emph{LLaMA-13B} \cite{Llama-2-13b} on several benchmarks, we did not experiment with these models.

In addition, our results suggest that Mixtral-8x7B-Instruct-v0.1~\cite{Mixtral} and GPT models outperform previous LLMs in understanding and following instructions as specified in the prompts in zero-shot learning scenarios.
Indeed, our experiments with \emph{gpt-3.5-turbo-0125}\cite{OpenAi}, \emph{Mixtral-8x7B-Instruct-v0.1}\cite{Mixtral}, and \emph{gpt-4-0125-preview}~\cite{OpenAi} show promising improvements when transitioning from sentence-level to paragraph-level passages. Table~\ref{tab:result} shows the improvements in the mean \emph{Accuracy} results obtained using these three models.


\begin{table}[h]
\centering
\caption{Mean Accuracy Results}
\label{tab:result}
\begin{tabular}{@{}lcc@{}}
\toprule
Model                            & Initial Accuracy (\%) & Improved Accuracy (\%) \\
    & (Sentence Level) & (Paragraph Level) \\

\midrule
GPT-3.5-Turbo-0125               & 30                    & 63                     \\
Mixtral-8x7B-Instruct-v0.1       & 33                    & 69                     \\
GPT-4-0125-Preview               & 41                    & 81                     \\ \bottomrule
\end{tabular}
\end{table}

Time constraints limited the scope of our evaluation. Nonetheless, initial findings through our zero-shot experiments support the superior analytical capabilities of these LLMs when broader textual contexts are integrated. We defer providing more detailed metrics for all the labels until we have evaluated the entire test data set with a balanced set of labels.

\textit{\textbf{RQ2.}}
The BERT-based classification approach by Ilyas et al.~\cite{ilyas2023multi} omits several compliance rules in its implementation and furthermore lacks annotated paragraph-level data. These issues preclude, at this stage, a conclusive comparison of our approach against that of Ilyas et al.'s to answer RQ2. To conduct a tentative comparison, considering that Ilyas et al. do not provide a public implementation of their approach, we have re-implemented it. This re-implementation, publicly available in our online repository~\cite{Code}, serves as our benchmark for comparison. The results obtained from this benchmark across two compliance rules with high prevalence yield an average \emph{F-score} of 67\%. In contrast, the lowest \emph{F-score} observed in our approach (using \emph{gpt-4-0125-preview} as the underlying LLM) exceeds 80\%, indicating major accuracy gains. Further analysis details can be found in our online material~\cite{RE2024Replication}.

\textit{\textbf{RQ3.}}
In our zero-shot learning experiment across a typical DPA, the total token count, total assistant token count, and the combined token count of both user and system are 25473, 769, and 24479, respectively. At the time of the experimentation (January 2024), the cost of processing our example DPA with GPT-3.5 Turbo and with GPT-4 Turbo using the OpenAI API was \$0.026 and \$0.27, respectively, subject to the OpenAI pricing at that time. The costs for an individual DPA are minimal (noting, of course, the order-of-magnitude disparity between GPT-3.5 Turbo and GPT-4 Turbo costs), and are expected to generate significant savings for lawyers, compliance experts, and consequently, businesses seeking legal services.
In interviews, our collaborating industry partner estimated that they dedicate two weeks and a team of three experts to each business. They foresee automation cutting mundane tasks and compliance costs, while emphasizing the indispensable role of human expertise alongside automation.
Lastly, in relation to execution time, we observe that our approach has an average execution time of approximately 0.7 seconds per paragraph. This suggests our implementation can handle more than 5000 paragraphs of textual legal content per hour. Considering that our approach can be run offline, we find the performance of our approach to be acceptable.


\subsection{Limitations and Validity Considerations} \label{sec:threats}

\textbf{Generalizability:} Our evaluation is preliminary. While the approach may work well for the specific datasets and compliance rules tested, we do not have enough evidence to conclude that it would generalize to other domains of law or regulatory frameworks. 
\textbf{Interpretability of Justifications:} While generative LLMs may provide justifications for their decisions, the interpretability of these justifications is crucial. They need to be clear and understandable to legal experts to be useful. Our evaluation has not yet examined the usefulness of the automated justifications provided. 
\textbf{Context Span:} While adding context that spans across sentences is beneficial, the context might also span multiple paragraphs, which could be considered as additional input for each prompt. Recent advancements in LLMs, e.g., the development of   Gemini 1.5 Pro~\cite{Gemini} with a context window of up to 1 million tokens, facilitate such extended context. However, the challenge of selecting the appropriate amount of context is still critical to avoid the ``needle in a haystack'' problem, where too much information can dilute and obscure relevant details. Our current evaluation suggests that paragraphs are a better context than sentences for compliance checking; however, the evaluation does not address the question of what the ``optimal'' context is for this task.

%
\textbf{Evolution:} Most legal texts are revised continuously over time.  Continuously updating the training data for LLMs is essential to ensure accuracy and relevance, distinguishing between up-to-date information and outdated ``hallucinations'' not aligned with current regulations. Our current work does not address the potential risks posed by deprecated legal texts and the difficulty of handling multiple versions of the same legal text.
\textbf{Content Chunking:} While investigating DPAs, we discovered through experiments that the optimal chunk of content, containing reasonable and necessary context for a sentence, is extracted when a line break followed by a heading is observed. Our current evaluation does not explore other regulatory artifacts, e.g., privacy policies, to determine the most suitable chunk for each sentence based on its context.
\textbf{Model Bias:} LLMs may inherit biases from their training data, which can affect the fairness and impartiality of compliance assessment. Our current evaluation does not offer insights into the perceived and actual severity of this risk.

\section{Future Research} \label{sec:conclusion}
The preliminary research presented in this paper aims to make a case for the need to reconsider current practices in legal compliance automation in light of recent advances in AI. Specifically, we argue that the substantially enhanced capacity of modern LLMs to handle context is likely to induce a major shift in our treatment of textual legal artifacts. This shift will involve transitioning from analyzing smaller contexts, such as individual sentences and phrases, to considering larger volumes of content, such as paragraphs and beyond, as context. We posit that the larger context will be able to provide the prerequisite knowledge, including cross-referenced legal materials, to create a self-contained basis for accurate automated decision-making regarding compliance and non-compliance.

Our future work will focus on four main aspects: (1)~enriching the DPA dataset~\cite{Amaral2023ML} with paragraph-level annotations; (2)~conducting comprehensive empirical evaluations to validate the effectiveness of paragraph-level context in increasing LLM accuracy, as defined in Section~\ref{sec:evaluation}; (3)~benchmarking against prior BERT-based approaches, e.g.,~\cite{ilyas2023multi}, to showcase comparative advantages; and (4)~seeking input from legal experts to review the outputs, particularly focusing on the explanation and justification provided by LLMs.

\section{Data Availability}\label{sec:package}
Our online repository is available at~\cite{RE2024Replication}. Specifically, the dataset can be found at~\cite{Data}, while the implementation of algorithms and evaluation scripts is provided at~\cite{Code}.

\section{Acknowledgment}
We gratefully acknowledge funding from MITACS under grant Number IT37879. The research was conducted while the first author was an intern at New Software (formerly Knowd) under the Vector Institute's FastLane program.

\makeatletter
\newcommand{\myfontsize}{\@setfontsize\myfontsize{8.25}{8.25}}
\makeatother

\let\oldthebibliography\thebibliography
\renewcommand{\thebibliography}[1]{%
  \oldthebibliography{#1}%
  \myfontsize 
  \setlength{\itemsep}{0pt}%
}

\bibliographystyle{IEEEtran}
\bibliography{ref}
\end{document}